\documentclass[aps,prl,reprint,twocolumn]{revtex4-2}

\usepackage{graphicx} 
\usepackage{bm}
\usepackage{amsmath}
\graphicspath{{./figures/}}
\usepackage{hyperref}
\hypersetup{colorlinks=true}
\usepackage{titlesec}
\usepackage[capitalise]{cleveref}
\usepackage{amssymb}
\usepackage{siunitx}
\usepackage{newtx}
\usepackage{chemformula}

\newcommand{\hc}{\mathrm{h.c.}}

\DeclareMathOperator{\sign}{sign}

\newcommand{\subfigref}[2]{Fig.~\hyperref[#1]{\ref*{#1}#2}}

\DeclareMathOperator{\Imm}{Im}

\newcommand{\ii}{\mathrm{i}}
\makeatletter \renewcommand\d[1]{\ensuremath{%
		\mathrm{d}#1\@ifnextchar\d{\!}{}}\;}
\makeatletter \newcommand\D[1]{\ensuremath{%
		\mathcal{D}#1\@ifnextchar\d{\!}{}}\;}
\makeatother

\newcommand{\Jsdzdefault}{\SI{-10}{meV}}
\newcommand{\Jsdz}{\SI{-10}{meV}}

\newcommand{\Jperp}{\SI{1}{meV}}

\newcommand{\Voffset}{\SI{5}{meV}}

\newcommand{\Vcrit}{\SI{1.35}{eV}}
\newcommand{\Ebinding}{\SI{0.25}{eV}}

\newcommand{\Deltamagnon}{\SI{1}{meV}}
\newcommand{\Gammaexciton}{\SI{0.2}{meV}}

\newcommand{\gilbert}{10^{-2}}
\newcommand{\gint}{\SI{5}{\micro eV \micro m^2}}

\newcommand{\Jsdzero}{\SI{-50}{meV}}
\newcommand{\latticeconstant}{\SI{1}{nm}}
\newcommand{\Egap}{\SI{1}{eV}}

\newcommand{\Jstiffness}{\SI{20}{meV \angstrom^2}}
\newcommand{\tunnel}{\SI{10}{meV}}
\newcommand{\Jeffa}{\SI{0.6}{meV}}
\newcommand{\Jeffo}{\SI{2.8}{meV}}
\newcommand{\Gammaa}{\SI{0.11}{\micro eV \mu m^2}}

\newcommand{\acousticgap}{\SI{0.4}{meV}}

\begin{document}
	\title{Bilinear magnon--exciton coupling in biased ferromagnetic electron--hole bilayers}
	\author{Pieter M. Gunnink}
	\email{pgunnink@uni-mainz.de}

	\affiliation{Institute of Physics, Johannes Gutenberg University Mainz, Staudingerweg 7, 55128 Mainz, Germany}
	
	\date{\today}
	\begin{abstract}
		The hybridization of magnons and excitons would combine magnetic and optical degrees of freedom in a single composite quasiparticle. Such a hybridization is however difficult to achieve, because of their inherent energy mismatch. 
		We propose that in biased bipolar ferromagnetic electron--hole bilayers the excitons and magnons can be brought into resonance, with the exciton energy lowered through the voltage bias to match the magnon energies. We demonstrate the linear hybridization of magnons and spin-flip excitons in this regime, starting from the microscopic exchange interactions between electrons and localized magnetic moments.
		We show that further increasing the gate voltage softens the hybrid magnon--exciton mode and realizes a magnon--exciton condensate, which manifests in both the magnon and exciton sectors and is associated with spin-superfluid transport. Ferromagnetic electron--hole bilayers therefore provide a new platform for the study of composite magnon--exciton quasiparticles and the realization of spinful condensates and associated spin superfluidity. 
	\end{abstract}
	\maketitle

	\paragraph{Introduction}
	Long-range magnetic order coexists with strongly bound excitons in two-dimensional magnetic semiconductors \cite{adakExcitonsVanWaals2026}, with the
	excitonic states and magnetic moments originating from the same electronic orbitals. Their coupling via intrinsic exchange interactions can therefore be strong, and excitons can exhibit a pronounced sensitivity to magnetic order: exhibiting magnetic control over the excitonic resonance \cite{wilsonInterlayerElectronicCoupling2021}, magnetic linear dichroism \cite{hwangboHighlyAnisotropicExcitons2021,dirnbergerSpincorrelatedExcitonPolaritons2022}, and an excitonic enhancement of magnetic circular dichroism \cite{wuPhysicalOriginGiant2019}. Beyond static magnetic-order effects, magnons can modulate exciton energies \cite{baeExcitoncoupledCoherentMagnons2022,dirnbergerMagnetoopticsVanWaals2023,diederichTunableInteractionExcitons2023}, induce exciton transport \cite{dirnbergerExcitonTransportDriven2026,iakovlevBoltzmannTransportTheory2026}, and drive exciton--exciton interactions \cite{dattaMagnonmediatedExcitonExciton2025}, while excitons can exert spin torques \cite{brennanExcitonicSpinTorque2026,varela-manjarresUltrafastOpticalExcitation2026}---demonstrating the unique dynamic tunability of magnetic order mediated by excitonic states \cite{brennanImportantElementsSpinExciton2024}.
	
	The magnon--exciton interactions that have been proposed and demonstrated thus far are between the exciton \emph{density} and magnons. This can be attributed to the inherent energy mismatch between magnons (meV) and excitons (eV), prohibiting linear hybridization. In electron--hole bilayers, however, exciton energies can be tuned down to the meV range. Such systems have attracted much attention recently because they offer a tunable platform for realizing excitonic insulators and excitonic Bose--Einstein condensates (BECs) \cite{xieElectricalReservoirsBilayer2018,maStronglyCorrelatedExcitonic2021,qiPerfectCoulombDrag2025,qiThermodynamicBehaviorCorrelated2023,zengElectricallyControlledTwodimensional2020,foglerHightemperatureSuperfluidityIndirect2014,qiTwocomponentExcitonCondensates2026,wuTheoryTwodimensionalSpatially2015,moonExcitonCondensateVan2025}.
	
\begin{figure}
		\centering
		\includegraphics[width=\columnwidth]{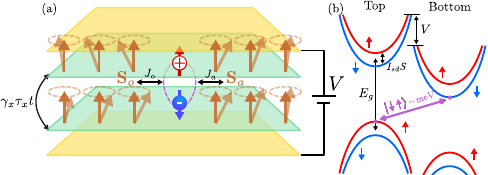}
		\caption{(a) A bipolar magnetic semiconducting bilayer with an applied voltage bias $V$ such that the lowest-energy exciton is the interlayer spin-flip electron--hole state illustrated in (b), whose energy can be tuned to the meV scale. A finite interlayer ($\gamma$) and interband ($\tau$) tunneling element $\tau_x\gamma_x t$ allows direct hybridization between excitons and acoustic and optical magnons.}
		\label{fig:setup}
	\end{figure}

	Inspired by these recent experimental advances, we consider in this work a bipolar magnetic semiconductor (BMS) bilayer \cite{chenRecentProgress2D2024, dengTwodimensionalBipolarFerromagnetic2021, liBipolarMagneticSemiconductors2012, liReviewBipolarMagnetic2022} with an applied gate voltage, as shown in \cref{fig:setup}, to realize a ferromagnetic (FM) electron--hole bilayer. The exciton energies can be tuned down to the meV range, while the electron states are coupled to the magnetic order through exchange interactions. Magnons and excitons are thus at comparable energies, allowing hybridization. From a microscopic electron--magnon model, we derive the bilinear coupling between excitons and magnons and demonstrate that it requires only finite interlayer and interband tunneling. The hybridization persists in the BEC phase, where the condensate is formed from magnon--exciton hybrids with corresponding magnon--exciton Bogoliubov excitations.

	\paragraph{Formalism}
	We consider a dual-gated bilayer magnetic semiconductor with localized magnetic moments, described by the Hamiltonian $H=H_m+H_e+H_{sd}+H_C$. We provide full details of the derivation in the Supplemental Material.\footnote{See the Supplemental Material for the microscopic derivation of the magnon and electron Hamiltonians, the exciton projection, the magnon--exciton couplings, and the Bogoliubov theory.} Taking into account intra- and interlayer exchange and an uni-axial anistropy, the magnon Hamiltonian,
	\begin{equation}
		H_m = \sum_{\bm k}\left(\phi_{a;\bm k}^\dagger \phi_{a;\bm k} \omega_{\bm k}^a +\phi_{o;\bm k}^\dagger \phi_{o;\bm k} \omega_{\bm k}^o\right). \label{eq:magnon-ham}
	\end{equation}
	gives acoustic and optical modes with dispersions
	\begin{equation}
		\omega^a_{\bm k} = \Delta_a + \rho_s k^2,\qquad \omega^o_{\bm k} = \Delta_o + \rho_s k^2.
	\end{equation}
	Here $\Delta_a$ is the acoustic magnon gap and $\Delta_o\equiv\Delta_a+2J'$ is the optical magnon gap, where $J'\equiv J^\perp S$ is the interlayer exchange energy and $\rho_s$ is the magnon spin stiffness. A full derivation is given in the End Matter.  In what follows, we set $S=1$ and choose the representative values $\rho_s=\Jstiffness$ \cite{shuSpinStiffnessChromiumbased2021}, $J'=\Jperp$, and $\Delta_a=\Deltamagnon$ \cite{chenTopologicalSpinExcitations2018}.
	
	The electron Hamiltonian is given by 
	\begin{equation}
		H_e=\sum_{\bm k}\sum_{\lambda\lambda'}
		d^\dagger_{\lambda;\bm k}[h_{\bm k}]_{\lambda\lambda'}d_{\lambda';\bm k},
	\end{equation}
	where
	\begin{equation}
		h_{\bm k}=\tau_z\left(\frac{E_g}{2}+\frac{k^2}{2m}\right)
		+\tau_x\gamma_xt+\gamma_z\frac V2-I_{sd}^{\gamma\tau}S\sigma_z.
	\end{equation}
	Here $\lambda=(\gamma,\tau,\sigma)$ combines the layer, conduction/valence-band, and spin indices. We assume equal electron masses, $m$, and allow for a finite interlayer tunneling element between the conduction and valence bands, parametrized by $t$. The static coupling to the local magnetic moments is described by an effective $sd$-like coupling, $I_{sd}^{\gamma\tau}$, which we allow to be layer- and band-dependent. For the BMS considered here, we have $I_{sd}^{\gamma\tau}<0$. The interlayer bias $V$ tunes the effective band gap, bringing the exciton energies into the meV range required for direct hybridization with magnons.

	We apply a voltage bias $V$ to bring the spin-down valence band of the top layer close in energy to the spin-up conduction band of the bottom layer [\subfigref{fig:setup}{(b)}], such that the interlayer spin-flip exciton is the lowest exciton state. Within this subspace, the effective electron Hamiltonian is
	\begin{equation}
		H_e=\sum_{\bm k}\left[c^\dagger_{\downarrow\bm k}\left(\frac{\Delta}{2}+\frac{k^2}{2m}\right)c_{\downarrow\bm k}-v^\dagger_{\uparrow\bm k}\left(\frac{\Delta}{2}+\frac{k^2}{2m}\right)v_{\uparrow\bm k}\right],
	\end{equation}
	where $c_{\downarrow\bm k}\approx d_{b,c,\downarrow;\bm k}$ and $v_{\uparrow\bm k}\approx d_{t,v,\uparrow;\bm k}$ to zeroth order in the tunneling and $\Delta=2\sqrt{t^2+(E_g/2-V/2)^2}-2|I_{sd}^{0}|S$ is the new band gap. The exciton formed by these bands is therefore a spin-flip exciton. Here $I_{sd}^{0/z}\equiv (I_{sd}^{bc}\pm I_{sd}^{tv})/2$. 
	
	The dynamic coupling to the magnetic moments is
	\begin{equation}
		H_{sd}=\frac{-1}{\sqrt{\mathcal N}}
		\sum_{\substack{\sigma\sigma'\\\tau\in\{c,v\}\\\gamma\in\{t,b\}}}
		\sum_{\bm q,\bm k}I_{sd}^{\gamma\tau}\bm S_{\gamma;\bm q}\cdot{}\\
		d_{\gamma\tau\sigma;\bm k+\bm q}^\dagger
		\bm\sigma_{\sigma\sigma'}d_{\gamma\tau\sigma';\bm k}.
	\end{equation}
	We project this interaction onto the spin-up valence band and spin-down conduction band, as shown in the End Matter,
	\begin{equation}
		H_{sd} =\frac{\sqrt{S}}{\sqrt{\mathcal N}}\sum_{\eta\in\{a,o\}}g_{\eta}\sum_{\bm q,\bm k} \left[
		c^\dagger_{\downarrow\bm k+\bm q}\phi_{\eta;\bm q}v_{\uparrow\bm k}+\hc
		\right],
	\end{equation}
	retaining only terms linear in the magnon operators, with
	\begin{equation}
		g_{\eta}\equiv-\frac{t}{\sqrt{t^2+(E_g/2-V/2)^2}}\times\begin{cases}
			I_{sd}^z,&\quad \eta=a,\\
			I_{sd}^0,&\quad \eta=o
		\end{cases}. \label{eq:geta}
	\end{equation}
	We have made here a long-wavelength and local approximation for the interaction vertex, neglecting its $\bm k$ dependence.

	Finally, we introduce an interband interaction, 
	\begin{multline}
		H_C=\frac{1}{2\mathcal V}\sum_{\sigma\sigma'}\sum_{\bm k,\bm k',\bm q}
		V(\bm k-\bm k')\,c^\dagger_{\sigma\,\bm q/2+\bm k}
		v^\dagger_{\sigma'\,\bm q/2-\bm k}\\
		\times v_{\sigma'\,\bm q/2-\bm k'}c_{\sigma\,\bm q/2+\bm k'},
	\end{multline}
	between the valence and conduction bands, which gives rise to bound electron--hole pairs, i.e., excitons. To capture the qualitative physics, we use the Coulomb interaction $V(\bm k)=e^2/(2\epsilon k)$, with $e$ the electron charge and $\epsilon=\epsilon_0\epsilon_r$, where $\epsilon_0$ is the vacuum permittivity and $\epsilon_r$ is the dimensionless dielectric constant. We neglect the layer separation $d$, which is valid when $d\ll a_{1s}$, with $a_{1s}$ the exciton radius. To model realistic conditions more accurately, we use the exciton binding energy $E_B$ as a free parameter and employ the Coulomb interaction only when calculating the magnon--exciton interaction.
	
	\paragraph{Magnon--exciton hybridization.} 
	To obtain a bilinear magnon--exciton coupling, we introduce excitons within a Matsubara path integral formalism \cite{noordmanVariationalFieldtheoreticalApproach2025}. We relegate the details to the SM \cite{Note1}, and state only the results here. We focus here for simplicity on the lowest-energy $1s$ excitonic state \cite{combescotExcitonsCooperPairs2015}, and obtain the effective action
	\begin{equation}
		\mathcal S = \sum_{\bm q}\int \d{\omega}\vec\Psi_{\bm q}^\dagger(\omega)
		{\setlength{\arraycolsep}{3pt}\begin{pmatrix}
				-G^{-1}_{X} & J_{\bm q}^{a} & J_{\bm q}^{o} \\
				J_{\bm q}^{a} & -G^{-1}_{a} & \\
				J_{\bm q}^{o} & & -G^{-1}_{o}
		\end{pmatrix}}\vec\Psi_{\bm q}(\omega), \label{eq:full-action}
	\end{equation}
	where $\vec\Psi_{\bm q}(\omega)\equiv (X_{\bm q} (\omega), \phi_{a;\bm q}(\omega), \phi_{o;\bm q}(\omega))^T$ contains the exciton, acoustic-magnon, and optical-magnon fields, and
	$G^{-1}_{X}=\omega + \ii \Gamma_X - \epsilon_{\bm q}$ and $G^{-1}_{a/o}=\omega + \ii \gamma_{a/o}-\omega_{\bm q}^{a/o}$ are the inverse exciton and magnon Green functions, respectively, with magnon damping $\gamma_{a/o}$ and exciton damping $\Gamma_X$. Here $\epsilon_{\bm q}=\Delta_X+q^2/(2M_X)$ is the exciton dispersion, $M_X=2m$ is the exciton effective mass and $\Delta_X\equiv\Delta-E_B$. Finally,
	\begin{equation}
		J_{\bm q}^{\eta}=-g_\eta F(1) \sqrt{\frac{\pi}{8}}\frac{l_{m}}{a_{1s}}\sqrt{\frac{E_B}{\Delta+q^2/2M_X-\omega}},
	\end{equation}
	is the magnon--exciton coupling for branch $\eta\in\{a,o\}$. Here $a_{1s}=1/\sqrt{2\mu E_B}$ is the exciton radius, $\mu=m/2$ is the reduced exciton mass, $l_m\equiv\sqrt{A_{\mathrm{uc}}S}$ is the magnon normalization length, with $A_{\mathrm{uc}}$ the unit-cell area, and $F(1)$ is a form factor, which is $F(1)\simeq0.886$ for $1s$-excitons.
	
	We can further write this in Hamiltonian form by making the on-shell approximation, $J_{\eta}\equiv J_{\bm q}^{\eta}|_{\omega\rightarrow\epsilon_{\bm q}}$, to obtain a coupled magnon--exciton Hamiltonian,
	\begin{equation}
		H_{Xm} =\sum_{\bm q}\vec\Psi_{\bm q}^\dagger
		\begin{pmatrix}
			\epsilon_{\bm q} & J_{a} & J_{o} \\
			J_{a} & \omega_{\bm q}^a & \\
			J_{o} & & \omega_{\bm q}^o
		\end{pmatrix}\vec\Psi_{\bm q} \label{eq:full-ham},
	\end{equation}
	where $\vec\Psi_{\bm q}\equiv (X_{\bm q},\phi_{a;\bm q},\phi_{o;\bm q})^T$ contains the exciton and magnon annihilation operators, and
	\begin{equation}
		J_{\eta} = -g_\eta F(1)\sqrt{\frac{\pi}{8}}\frac{l_{m}}{a_{1s}}
	\end{equation} 
	is the effective magnon--exciton coupling.

	\begin{figure}
		\centering
		\makebox[\columnwidth][c]{%
			\begin{tikzpicture}
				\node at (0,2) {\includegraphics{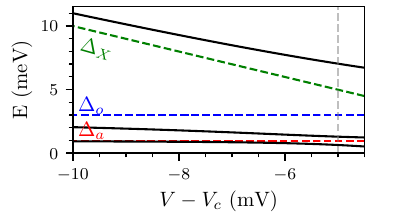}};
				\node at (0,-1.9) {\includegraphics{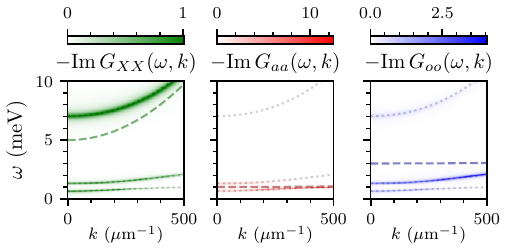}};
				\node at (-4,3.8) {(a)};
				\node at (-4,0) {(b)};
			\end{tikzpicture}%
		}
		\caption{(a) The $\bm k=0$ hybridization of the exciton ($\Delta_X$) and the acoustic and optical magnons (with gaps $\Delta_{a/o}$) as the gate voltage $V$ is increased. $V_c$ is the critical voltage at which $\Delta_X=0$ . The gaps in the absence of coupling are shown by dashed lines. The effective magnon--exciton couplings are $J_a=\Jeffa$ and $J_o=\Jeffo$. (b) The exciton and magnon spectral density, $-\Imm[G_{\alpha\alpha}(\omega,k)]$, for $V-V_c=\Voffset$ (indicated by a vertical dashed line in (a)). Dotted lines show the energies obtained from the Hamiltonian approximation, and the bold dashed lines show the bare exciton and magnon energies.}\label{fig:hybrid}
	\end{figure}
	
	We now study the magnon--exciton hybridization as the exciton energy is lowered through the gate voltage $V$. We choose a moderate tunneling strength $t=\tunnel$ \cite{sorianoInterplayInterlayerExchange2019}, and $E_g=\Egap$, $a=\latticeconstant$, $E_B=\Ebinding$ \cite{azhikodanAnomalousInterlayerExciton2016,kambanInterlayerExcitonsVan2020}, $I_{sd}^0=\Jsdzero$ \cite{kimEvolutionInterlayerIntralayer2019}, $\Gamma_X=\Gammaexciton$, $\gamma_a=\gilbert\Delta_a$, and $\gamma_o=\gilbert\Delta_o$ \cite{grzeszczykStronglyCorrelatedExcitonMagnetization2023}. We set $I_{sd}^z=\Jsdz$ unless stated otherwise.
	
	We show the resulting zero-momentum energies in \subfigref{fig:hybrid}{(a)}, obtained from diagonalization of the Hamiltonian, \cref{eq:full-ham}. Here $V_c=E_g+\sqrt{(E_B+I_{sd}^0S)^2-4t^2}\approx \Vcrit$ is the critical voltage at which the zero-momentum exciton energy reaches zero.
	As the voltage is increased, the exciton energy decreases, which allows hybridization with the magnon branches and leads to a three-level avoided crossing. 
	This coupling persists at finite momentum, as we demonstrate by showing the exciton and magnon spectral density, $-\Imm[G_{\alpha\alpha}(\omega,k)]$, in \subfigref{fig:hybrid}{(b)} for $V-V_c=\Voffset$. Overlaid are the solutions from the Hamiltonian approximation (dotted lines) and the bare, uncoupled energies (bold dashed lines). We observe that the Hamiltonian approximation is excellent and matches the resonances in the spectral density. Owing to the strongly dispersive character of the exciton compared with that of the magnons, the magnon dispersion displays a dip as the exciton dispersion approaches it from above, which could be measured with Brillouin light scattering \cite{sampaioDzyaloshinskiiMoriyaInteraction2025}.
	Furthermore, the spectral weights of the acoustic and optical magnons are redistributed across the branches, even at $k=0$, which could be visible in magneto-Raman spectroscopy \cite{cenkerDirectObservationTwodimensional2021}.
	We conclude that excitons can display sizable hybridization with acoustic and optical magnons.

	\paragraph{Bose--Einstein condensation.} If the excitons are lowered further in energy, an exciton Bose--Einstein condensate is formed. In this scenario, the coupling to the magnon subsystem naturally generates multicomponent order \cite{kawaguchiSpinorBoseEinstein2012,qiTwocomponentExcitonCondensates2026}, which we will study here.
	
	For the formation of a BEC, higher-order interactions are key to stabilizing the condensate \cite{pethickBoseEinsteinCondensation2006,griffinBoseCondensedGasesFinite2009}. Within the exciton sector, this is the repulsive exciton--exciton interaction. In the FM bilayer considered here, there is also a coupling between the magnon and exciton densities, as we show in the SM \cite{Note1}. The Euclidean-time action of the three-boson system is thus
	
	\begin{align}
		\mathcal S=\int_x\Bigg\{&
		\bar X\left[
		\partial_\tau+\Delta_X-\frac{\nabla^2}{2M_X}
		\right]X+\frac12 g n_X^2 +\Gamma_0n_X\left(n_a+n_o\right)
		\nonumber\\
		&\hspace{-4em}+\sum_{\eta\in\{a,o\}}\bar\phi_\eta\left(
		\partial_\tau+\Delta_\eta-\rho_s\nabla^2
		\right)\phi_\eta+\left(\bar XJ_\eta\phi_\eta+\hc\right), \label{eq:SBEC}
	\end{align}
	where $g>0$ is the repulsive exciton--exciton interaction strength. The magnon--exciton density--density coupling is
	\begin{equation}
		\Gamma_{0}\equiv -\sqrt{2}A_{\mathrm{uc}}\frac{E_B^2}{\Delta^2}
		I_{sd}^0,
	\end{equation}
	which is thus repulsive, since $I_{sd}^0<0$ in a BMS. We used the Hamiltonian approximation for the bilinear coupling, and $\int_x\equiv\int\d{\tau}\d{\bm x}$. We choose $g=\gint$ \cite{qiTwocomponentExcitonCondensates2026}, and for $I_{sd}^0=\Jsdzero$ have $\Gamma_0=\Gammaa$.
	
	\begin{figure*}
		\begin{tikzpicture}
			\node at (-0.66\columnwidth,0) {\includegraphics[]{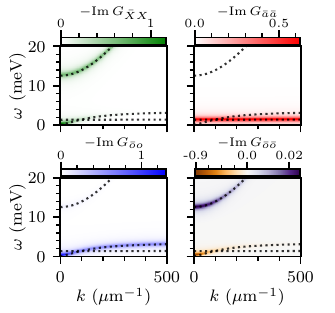}};
			\node at (0,0) {\includegraphics{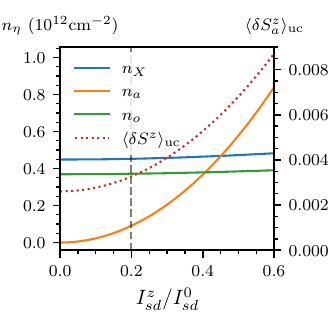}};
			\node at (0.66\columnwidth,0) {\includegraphics[]{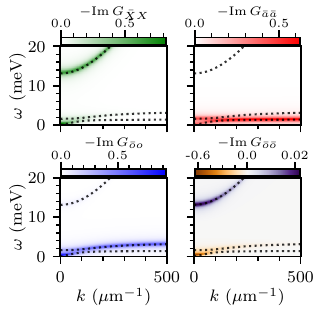}};
		\end{tikzpicture}
		\caption{(a) The spectral density, $-\Imm[G_{\alpha\beta}]$, in the condensed phase, with $I_{sd}^z=0$, such that $J_a=0$ and thus $n_a$ is zero. The optical magnon couples to the Goldstone mode of the condensate, while the finite repulsive coupling $\Gamma_0$ induces a positive shift, $\Gamma_0 n_X=\acousticgap$, of both the optical and acoustic modes. Coupling to the exciton sector induces two-mode squeezing of the optical mode, such that $-\Imm[G_{\bar o\bar o}]\neq0$. (b) The exciton and magnon condensate densities $n_X,n_a,n_o$ (left axis) as a function of $I_{sd}^z$, with $I_{sd}^0=\SI{50}{meV}$. The right axis shows the condensate-induced spin polarization per unit cell. (c) The spectral density with $I_{sd}^z=\Jsdzdefault$, such that both $J_{a,o}\neq0$ and thus the Goldstone mode hybridizes with the acoustic and optical magnon modes.  An additional broadening of \(\SI{0.5}{meV}\) is included for visual clarity.}
		\label{fig:BEC}
	\end{figure*}

	We expand the exciton and magnon fields, $X=X_0+X'$, $\phi_{a/o}=\phi_{a/o;0}+\phi_{a/o}'$, around a homogeneous and time-independent condensate to find Gross--Pitaevskii equations,
	\begin{align}
		\left[\Delta_X + gn_X+\Gamma_0 (n_a+n_o)\right]X_0+J_a\phi_{a;0}+J_o\phi_{o;0}&=0, \\
		\left[\Delta_a+\Gamma_0 n_X\right]\phi_{a;0}+J_aX_0&=0, \\
		\left[\Delta_o+\Gamma_0 n_X\right]\phi_{o;0}+J_oX_0&=0,
	\end{align}
	where $n_X\equiv |X_0|^2$ and $n_{a/o}\equiv |\phi_{a/o;0}|^2$ are the exciton and magnon condensate densities, respectively.
	It is instructive to first solve the case where $I_{sd}^z=0$, such that $J_a=0$. We then have $n_a=0$ and  
	\begin{equation}
		n_X=\frac{J_o^2/\Delta_o-\Delta_X}{g} + O\left(\frac{\Gamma_0}{g}\right);\quad n_o=\frac{J_o^2}{\Delta_o^2}n_X+ O\left(\frac{\Gamma_0}{g}\right).
	\end{equation}
	This corresponds to a condensate formed by the hybrid optical magnon--exciton mode, with a renormalized gap given by $J_o^2/\Delta_o-\Delta_X$, and the coupling $J_o$ induces a finite optical magnon density. The corrections up to first order in $\Gamma_0/g$ are given in the End Matter.
	
	In the general case, where all couplings are nonzero, we find
	\begin{equation}
		n_{a/o} = \frac{J_{a/o}^2}{(\Delta_{a/o}+\Gamma_{0} n_X)^2}n_X, \label{eq:nao}
	\end{equation}
	while the exciton condensate is given by the physical root of
	\begin{equation}
		\Delta_X + g n_X-\sum_{\eta\in\{a,o\}}\frac{ J_\eta^2\Delta_\eta}{(\Delta_\eta+\Gamma_0 n_X)^2} = 0. \label{eq:root-X0}
	\end{equation} 
	We solve this equation numerically and identify the physical root through the following properties: (i) the relative phase of the acoustic/optical magnon and exciton condensates is determined by $J_\eta$, such that $\sign{X_0/\phi_{\eta;0}}=-\sign{J_\eta}$; (ii) the second derivative of the free energy is positive.
	
	We show the resulting condensate densities in \subfigref{fig:BEC}{(b)} as a function of $I^z_{sd}$. On the right axis, we show the total spin polarization per unit cell, $\langle \delta S^z\rangle\equiv (n_{a}+n_o)A_{\mathrm{uc}}/\sqrt{2}$, which could serve as an experimental probe of condensate formation. We first observe that even at $I_{sd}^z=0$, the optical magnon density is comparable in magnitude to the exciton density. At finite $I_{sd}^z$, the acoustic magnons also condense. Since the acoustic magnon gap is smaller than the optical magnon gap, the acoustic magnon density can exceed the optical magnon density even if $|J_a|<|J_o|$; see \cref{eq:nao}. We finally note that the phase diagram is identical for negative $I_{sd}^z/I_{sd}^0$.
	
	We also obtain the Bogoliubov Green function from the part of the action quadratic in fluctuations,
	\begin{equation}
		\mathcal S=-\sum_{\bm k}\sum_n\vec\psi^\dagger(k)\hat G_B^{-1}(k)\vec\psi(k), \label{eq:full-bog}
	\end{equation}
	where $\vec\psi(k)=\left(X'(k),\bar X'(k),\phi'_a(k),\bar \phi'_a(k),\phi'_o(k),\bar \phi'_o(k)\right)^T$ is the Nambu spinor. Here 
	\begin{equation}
		-\hat G_B^{-1}(k)\equiv\begin{pmatrix}
			-\hat G_X^{-1}(k) & \hat T_a(k) & \hat T_o(k)	\\
			\hat T_a(k) & -\hat G_a^{-1}(k) & \\
			\hat T_o(k) & & -\hat G_o^{-1}(k)
		\end{pmatrix}, \label{eq:GBinv}
	\end{equation} 
	with $\hat G_X^{-1}$, $\hat G_{a/o}^{-1}$, and $\hat T_{a/o}$ the Bogoliubov exciton Green function, the acoustic/optical magnon Green functions, and their coupling matrices to the exciton fields, respectively, as given in the End Matter. Here $k=(\bm k,\ii\Omega_n)$.
	
	We numerically invert the Bogoliubov Green function and show the resulting spectral density, $-\Imm[G_{\alpha\beta}]$, in \subfigref{fig:BEC}{(a,c)}. The dotted lines are the eigenenergies obtained from a paraunitary diagonalization \cite{colpaDiagonalizationQuadraticBoson1978} of the corresponding Bogoliubov Hamiltonian \cite{Note1}. Focusing first on the $I_{sd}^z=0$ case [\subfigref{fig:BEC}{(a)}], we observe a linearly dispersing Goldstone mode associated with spontaneous breaking of the $U(1)$ symmetry, which hybridizes with the optical magnon branch. The Goldstone mode has weight in the $-\Imm[G_{\bar o o}]$ sector, and is thus a transverse-spin phase mode, associated with spin superfluidity.
	
	In the long-wavelength limit and for $J_a=\Gamma_0=0$, we can find the two energies of the optical magnon--exciton subsystem,
	\begin{equation}
		E_+=\frac{1}{\Delta_o}\sqrt{(J_o^2+\Delta_o^2)^2+2J_o^2\Delta_o gn_X};
		\qquad E_-=\frac{\Delta_o}{E_+}ck,
	\end{equation}
	where $c=\sqrt{{g n_X}/{M_X}}$ is the bare sound velocity \cite{stoofUltracoldQuantumFields2009}, which is renormalized by $\Delta_o/{E_+}$. The gap of the upper mode is also enhanced compared to a conventional two-level splitting, bringing the upper branch into the \SI{20}{meV} range. 
	
	Furthermore, even though the acoustic magnon density $n_a$ is zero, the acoustic magnon fluctuations still couple to the exciton density through the magnon--exciton density--density interaction $\Gamma_0$. This renormalizes the magnon gaps, $\Delta_{a/o}\rightarrow\Delta_{a/o}+\Gamma_0 n_X$, which for the parameters chosen here is a positive energy shift of $\Gamma_0 n_X=\acousticgap$. Finally, we also show the spectral density $-\Imm[G_{\bar o\bar o}]$, i.e., the optical magnon anomalous spectral function, which describes the two-mode squeezing correlations. This correlator demonstrates that although there are no anomalous couplings within the magnon sector, the coupling to the exciton sector induces squeezing.

	In \subfigref{fig:BEC}{(c)}, we show the general case, where $I_{sd}^z=\Jsdzdefault$ and thus all coupling elements are nonzero. This is reflected in the Bogoliubov spectrum, where we observe a three-level repulsion: the linearly dispersing sound mode hybridizes with both magnon modes. Both magnon gaps are also shifted.

	\paragraph{Conclusion and discussion.} We have demonstrated bilinear coupling between magnons and spin-flip excitons in a bipolar ferromagnetic electron--hole bilayer. A voltage bias brings the exciton and magnon energies into resonance, with exchange interactions between electrons and localized magnetic moments generating a sizable coupling. Lowering the bias further generates a magnon--exciton condensate and associated spin superfluidity, with magnon and exciton signatures.
	
	We consider here the exciton energy to be negative to induce a condensate \cite{kanekoNewEraExcitonic2025,maStronglyCorrelatedExcitonic2021}, but a more experimentally relevant setup might be a nonequilibrium condensate maintained by injecting electrons and holes via conducting leads \cite{xieElectricalReservoirsBilayer2018,zengElectricallyControlledTwodimensional2020,wangExcitonicTopologicalOrder2023,wangEvidenceHightemperatureExciton2019}. This setup would not directly realize magnon--exciton coupling in the uncondensed regime, because the exciton energies do not reach the \si{meV} range, but we still expect coupling between the collective condensate modes and magnons. The dynamical nature of the condensate would furthermore lead to an ac Josephson effect \cite{sunDynamicalExcitonCondensates2024}. An accurate treatment would require a full nonequilibrium theory \cite{sunDynamicalExcitonCondensates2024,zengKeldyshFieldTheory2024}, which is beyond the scope of this work.
	
	We have considered here a bipolar ferromagnetic semiconductor \cite{chenRecentProgress2D2024, dengTwodimensionalBipolarFerromagnetic2021, liBipolarMagneticSemiconductors2012, liReviewBipolarMagnetic2022}, where spin-flip excitons are the lowest-energy excitations. Alternative pathways towards bilinear magnon--exciton coupling include biased  A-type antiferromagnetic electron--hole bilayers \cite{lvElectricFieldTunableMagnetism2021,yaoSwitchingSpinPolarization2025,guoSpinPolarizedAntiferromagnetsSpintronics2025}, or spin-orbit coupling enabling the coupling between spin-zero excitons and magnons.

	\paragraph{Acknowledgments}
	P.\,M.\,G. is funded by the European Union through an MSCA Postdoctoral Fellowship (Project No. 101145915). P.\,M.\,G. thanks Jairo Sinova and Alexander Mook for valuable feedback on this manuscript.
	\bibliography{coupling}
	\newpage\section*{End Matter}
	\paragraph{Magnon dynamics. } We assume a minimal Heisenberg exchange model for the ferromagnetic bilayer,
	\begin{multline}
		H_m=-\sum_{\gamma\in\{t,b\}}\biggl[\sum_{\langle ij\rangle}J^\parallel\bm S_{\gamma,i}\cdot\bm S_{\gamma,j}+\sum_iJ^\perp\bm S_{\gamma,i}\cdot\bm S_{\bar\gamma,i}\\
		+\frac{K}{2}\sum_i(S_{\gamma,i}^z)^2\biggr],
	\end{multline}
	with intralayer exchange $J^\parallel>0$, interlayer exchange $J^\perp>0$, and uniaxial anisotropy $K$ along the $z$ axis. After a Holstein--Primakoff transformation, $S^+_{\gamma,i}\simeq\sqrt{2S}\phi_{\gamma,i}$, and a Fourier transformation, $\phi_{\gamma;i}=\frac{1}{\sqrt{\mathcal N}}\sum_{\bm k}e^{\ii\bm k\cdot\bm r_{\gamma;i}}\phi_{\gamma;\bm k}$, the long-wavelength quadratic magnon Hamiltonian is
	\begin{equation}
		H_m=\sum_{\gamma}\sum_{\bm k}\left[(\Delta_a+\rho_s k^2+J')\phi_{\gamma;\bm k}^\dagger\phi_{\gamma;\bm k}-J'\phi_{\gamma;\bm k}^\dagger\phi_{\bar\gamma;\bm k}\right],
	\end{equation}
	where $\Delta_a\equiv SK$, $\rho_s\equiv J^\parallel Sa^2$, and $J'\equiv J^\perp S$.
	This decomposes into the acoustic and optical modes,
	\begin{equation}
		\phi_{a;\bm k}=\frac{\phi_{t;\bm k}+\phi_{b;\bm k}}{\sqrt{2}};\qquad
		\phi_{o;\bm k}=\frac{\phi_{t;\bm k}-\phi_{b;\bm k}}{\sqrt{2}},
	\end{equation}
	which yields the magnon Hamiltonian in \cref{eq:magnon-ham}.
	\paragraph{Magnon-electron coupling. }
	We restrict ourselves to the basis spanned by the bands $\gamma=b,\tau=c$ and $\gamma=t,\tau=v$. The corresponding low-energy conduction- and valence-band Hamiltonian is
	\begin{equation}
		H_e=\sum_{\sigma,\bm k}\Biggl[
		c^\dagger_{\sigma\bm k}E_{\sigma\bm k}c_{\sigma\bm k}
		-v^\dagger_{\sigma\bm k}E_{\sigma\bm k}v_{\sigma\bm k}
		\Biggr],
	\end{equation}
	where $E_{\sigma\bm k}\equiv\sqrt{t^2+(E_g/2-V/2-\sigma I_{sd}^z S)^2}+k^2/2m$ and we neglect corrections to the effective mass, which are of the order $t^2/(E_g/2-V/2)^2$. Furthermore,
	\begin{align}
		c_{\sigma\bm k}&=
		\cos\frac{r_{\sigma\bm k}}{2}d_{b,c,\sigma;\bm k}
		+\sin\frac{r_{\sigma\bm k}}{2}d_{t,v,\sigma;\bm k},\\
		v_{\sigma\bm k}&=
		-\sin\frac{r_{\sigma\bm k}}{2}d_{b,c,\sigma;\bm k}
		+\cos\frac{r_{\sigma\bm k}}{2}d_{t,v,\sigma;\bm k},
	\end{align}
	with 
	\begin{equation}
		\tan r_{\sigma\bm k} = \frac{t}{\xi_{\bm k} + E_g/2-V/2-\sigma I_{sd}^z}.
	\end{equation}
	
	We now restrict ourselves to the subspace spanned by the highest valence and lowest conduction bands, i.e.,
	\begin{equation}
		H_e=\sum_{\bm k}\left[
		c^\dagger_{\downarrow\bm k}\left(\frac{\Delta}{2}+\frac{k^2}{2m}\right)c_{\downarrow\bm k}
		-v^\dagger_{\uparrow\bm k}\left(\frac{\Delta}{2}+\frac{k^2}{2m}\right)v_{\uparrow\bm k}
		\right],
	\end{equation}
	where 
	\begin{align}
		\Delta&\equiv E_{\uparrow}+E_{\downarrow}-2|I_{sd}^0|S\\
		&\approx2\sqrt{t^2+(E_g/2-V/2)^2}-2|I_{sd}^0|S
	\end{align}
	where we neglected corrections of the band gap due to  $I_{sd}^zS$, which are of order $(I_{sd}^zS)^2/(E_g/2-V/2)^2$. 
	 
	Within this subspace, tunneling modifies the interaction with the magnetic sector. We focus on possible direct hybridization with magnons and therefore retain only interband terms proportional to $S^{\pm}_{\gamma;\bm q}$. We then have
	\begin{equation}
		H_{sd} =\frac{1}{\sqrt{\mathcal N}}\sum_{\bm q,\bm k} \left[
		c^\dagger_{\downarrow\bm k + \bm q} W[\bm S_{\bm q}] v_{\uparrow\bm k} + \hc
		\right].
	\end{equation}
	Here
	\begin{equation}
		W[\bm S_{\bm q}]=-\frac{1}{\sqrt{2}}\frac{t}{\sqrt{t^2+(E_g/2-V/2)^2}}
		\left(I_{sd}^zS^+_{a;\bm q}+I_{sd}^0S^+_{o;\bm q}\right).
	\end{equation}
	We have defined $\sqrt{2}\bm S_{a/o;\bm q}\equiv\bm S_{t;\bm q}\pm\bm S_{b;\bm q}$ as the acoustic and optical modes. We emphasize that even if the top and bottom layers are made of the same ferromagnetic semiconductor, different orbital compositions of the valence and conduction bands can generate a sizable difference between $I_{sd}^{bc}$ and $I_{sd}^{tv}$ \cite{grzeszczykStronglyCorrelatedExcitonMagnetization2023}.
	Consistent with the local approximation, we neglect the $\bm k$-dependence of the interaction vertex, $W_{\bm k}[\bm S_{\bm q}]\approx W[\bm S_{\bm q}]$. This amounts to assuming $k^2/2m\ll|E_g/2-V/2|$ for the relevant internal exciton momenta $\bm k$. After the Holstein--Primakoff transformation, this yields the coupling shown in the main text [\cref{eq:geta}].
	
	\paragraph{Magnon--exciton density--density coupling.} We discuss here the main characteristics of the magnon--exciton density--density coupling, which we derive in full detail in the SM \cite{Note1}. This coupling originates from the longitudinal exchange interaction, which projected onto the two low-energy bands is
	\begin{equation}
		H_{sd}^z =\frac{1}{\sqrt{\mathcal N}}\sum_{\bm q,\bm k} \left[
		c^\dagger_{\downarrow\bm k + \bm q} W^z_c[\bm S_{\bm q}] c_{\downarrow\bm k} + v^\dagger_{\uparrow\bm k + \bm q} W^z_v[\bm S_{\bm q}] v_{\uparrow\bm k}
		\right],
	\end{equation}
	with
	\begin{align}
		W^z_c[\bm S_{\bm q}]&=\frac{I_{sd}^{bc}}{\sqrt{2}}(S^z_{a;\bm q}-S^z_{o;\bm q}),\\
		W^z_v[\bm S_{\bm q}]&=-\frac{I_{sd}^{tv}}{\sqrt{2}}(S^z_{a;\bm q}+S^z_{o;\bm q}).
	\end{align}
	We perform the full calculation in the SM, where we focus only on the equal-space and equal-time interaction, such that we have the action, in momentum and Matsubara-frequency space,
	\begin{equation}
		\mathcal S_z=
		\sum_{\eta\in\{a,o\}}
		\sum_{n,m}
		\sum_{\bm p,\bm q}
		\tilde\Gamma_\eta\,
		S^z_{\eta;\bm p}(\ii\Omega_m)
		X_{\bm q}(\ii\Omega_n)
		\bar X_{\bm q+\bm p}(\ii\Omega_{n+m}).
	\end{equation}
	where $\tilde\Gamma_\eta$ is given in the SM \cite{Note1}. Upon Fourier transforming to real space and Euclidean time, and 
	transforming $S^z_{a/o}$ to magnon operators, $\sqrt{2}S^z_{a}=2S-n_a-n_o$ and $\sqrt{2}S^z_{o}=-\bar\phi_o\phi_a -\bar\phi_a\phi_o$,
	we obtain a local-in-spacetime interaction as used in \cref{eq:SBEC}. We also induce hybridization between the acoustic and optical magnons, $n_X\bar\phi_o\phi_a + \hc$. These terms we neglect in our subsequent analysis of the BEC, because they are of the size $I_{sd}^z$ and thus typically smaller than the magnon--exciton density--density coupling. Furthermore, we do not expect these hybridizations to significantly influence the resulting BEC, only providing a small mixing between the acoustic and optical magnons, which will grow as the exciton density increases.

	\paragraph{$\Gamma_0$-corrections to the condensate.}The condensate for $I_{sd}^z=0$, such that $J_a=0$, can be solved in the limit of small $\Gamma_0$,
	\begin{align}
		n_X&=\frac{J_o^2/\Delta_o-\Delta_X}{g}\left[1-\frac{2J_o^2}{\Delta_o^2}\frac{\Gamma_0}{g}\right]+O(\Gamma_0^2/g^2),
		\\
		n_a&=0\\
		n_o&=\frac{J_o^2}{\Delta_o^2}\left[1-\frac{2(J_o^2/\Delta_o-\Delta_X)}{\Delta_o}\frac{\Gamma_0}{g}\right]n_X+O(\Gamma_0^2/g^2).
	\end{align}
	The exciton--magnon density--density interaction thus introduces repulsive corrections to the magnon and exciton densities, which are small in $\Gamma_0/g$, which for our parameters, $g=\gint$ and $\Gamma_0=\Gammaa$ is indeed small.
	\paragraph{Bogoliubov Green functions.}
	The exciton Green function, acoustic/optical magnon Green functions, and coupling matrices appearing in \cref{eq:GBinv} follow from expanding the action to quadratic order in fluctuations \cite{stoofUltracoldQuantumFields2009} and are given by \cite{Note1}
	\begin{align*}
		-\hat G_X^{-1}(k)&=-\ii\Omega_n\hat\sigma_z+
		\Bigl(\Delta_X+\frac{k^2}{2M_X}+2gn_X\\
		&\hspace{5em}+\Gamma_0(n_a+n_o)\Bigr)\hat\sigma_0
		+gn_X\hat\sigma_x,\\
		-\hat G^{-1}_{a/o}(k)&=-\ii\Omega_n\hat\sigma_z+
		\left(\omega^{a/o}_{\bm k}+\Gamma_{0}n_X\right)\hat\sigma_0,\\
		\hat T_{a/o}(k)&=J_{a/o}\hat\sigma_0+
		\Gamma_{0}\sqrt{n_Xn_{a/o}}
		\left(\hat\sigma_0+\hat\sigma_x\right).
	\end{align*}
	
\end{document}


\title{Supplemental Material for ``Bilinear magnon--exciton coupling in biased ferromagnetic electron--hole bilayers''}
	\author{Pieter M. Gunnink}
	\email{pgunnink@uni-mainz.de}
	\affiliation{Institute of Physics, Johannes Gutenberg University Mainz, Staudingerweg 7, 55128 Mainz, Germany}
	\date{\today}
	\maketitle
	
	\section{System}
	
	We describe the coupled magnon--electron system in the Matsubara path-integral formalism \cite{stoofUltracoldQuantumFields2009}. Its partition function is
	\begin{equation}
		\mathcal Z=\int\DD{\psi,\bar\psi,\phi,\bar\phi}
		e^{-\mathcal S_e[\psi,\bar\psi]-\mathcal S_C[\psi,\bar\psi]-\mathcal S_m[\phi,\bar\phi]}.
	\end{equation}
	Here $(\bar\psi,\psi)$ and $(\bar\phi,\phi)$ are the electron and magnon fields, respectively. We use $x=(\bm x,\tau)$ and
	$\int_x\equiv\int\d{\bm x}\d{\tau}$, where the spatial integral is over the system volume $\mathcal V$ and the imaginary time integral $\tau$ from  $0$ to $\beta\equiv1/T$. We set $\hbar=k_B=1$.
	
	The magnon action is
	\begin{equation}
		\mathcal S_m=-\sum_\eta\int_{x,x'}
		\bar\phi_\eta(x)G^{-1}_{0;\eta}(x,x')\phi_\eta(x'),
	\end{equation}
	where
	\begin{equation}
		-G^{-1}_{0;\eta}(x,x')=
		\left[\partial_\tau+H_\eta(-\ii\nabla)\right]\delta(x-x'),
	\end{equation}
	where $H_\eta(-\ii\nabla)$ is the long-wavelength Hamiltonian of magnon branch $\eta$.
	
	We restrict the electronic sector to conduction ($\tau=c$) and valence ($\tau=v$) bands. Allowing for a magnon self-energy that is not diagonal in this band basis gives
	\begin{equation}
		\mathcal S_e=-\sum_{\sigma\sigma'}\sum_{\tau,\tau'\in\{c,v\}}\int_{x,x'}
		\bar\psi_{\tau\sigma}(x)
		\left[G^{-1}_{0;\tau\sigma\sigma'}(x,x')\delta_{\tau\tau'}
		-\Sigma^m_{\tau\tau';\sigma\sigma'}(x,x')\right]\psi_{\tau'\sigma'}(x'),
	\end{equation}
	with
	\begin{equation}
		-G^{-1}_{0;\tau\sigma\sigma'}(x,x')=
		\left[\partial_\tau+\zeta^\tau_\sigma(-\ii\nabla)\right]
		\delta_{\sigma\sigma'}\delta(x-x').
	\end{equation}
	Here $\zeta^\tau_\sigma(-\ii\nabla)=\varepsilon^\tau_\sigma(-\ii\nabla)-\mu_\tau$ is the electron dispersion $\varepsilon^\tau_\sigma(-\ii\nabla)$ measured relative to the chemical potential. We assume a local magnon self-energy,
	\begin{equation}
		\Sigma^m_{\tau\tau';\sigma\sigma'}(x,x')=
		\Sigma^m_{\tau\tau';\sigma\sigma'}(x)\delta(x-x'),
	\end{equation}
	which will couple the electron density to the magnons.
	
	Finally, we include the interband interaction
	\begin{equation}
		\mathcal S_C=\frac12\sum_\tau\sum_{\sigma\sigma'}\int_{x,x'}
		\bar\psi_{\tau\sigma}(x)\bar\psi_{\bar\tau\sigma'}(x')V(x-x')\psi_{\bar\tau\sigma'}(x')\psi_{\tau\sigma}(x),
	\end{equation}
	with the instantaneous interaction $V(x-x')=V(\bm x-\bm x')\delta(\tau-\tau')$. Intraband interactions would generate additional corrections to the exciton Green function and are not considered here \cite{noordmanVariationalFieldtheoreticalApproach2025}.
	
	\section{Exciton fields}
	
	We introduce a bilocal polarization field $\mathcal P_{\sigma\sigma'}(x,x')$ through a Hubbard--Stratonovich transformation, following Ref.~\cite{noordmanVariationalFieldtheoreticalApproach2025}, choosing
	\begin{equation}
		\langle\mathcal P_{\sigma\sigma'}(x,x')\rangle=
		\langle\bar\psi_{c\sigma}(\bm x,\tau)\psi_{v\sigma'}(\bm x',\tau)\rangle
		\delta(\tau-\tau').
	\end{equation}
	The corresponding identity is
	\begin{equation}
		1=\int\DD{\mathcal P,\bar{\mathcal P}}\exp\Biggl\{-\int_{x,x'}
		\left[\bar{\mathcal P}_{\sigma'\sigma}(x',x)-\bar\psi_{c\sigma}(x)\psi_{v\sigma'}(x')\right]V(x-x')
		\left[\mathcal P_{\sigma\sigma'}(x,x')-\bar\psi_{v\sigma'}(x')\psi_{c\sigma}(x)\right]\Biggr\}.
	\end{equation}
	After this transformation, the interaction contribution becomes
	\begin{equation}
		\mathcal S_C=\int_{x,x'}\bar{\mathcal P}_{\sigma\sigma'}(x,x')V(x-x')
		\mathcal P_{\sigma'\sigma}(x',x)
		-\int_{x,x'}\bar\psi_{c\sigma}(x)\mathcal P_{\sigma\sigma'}(x,x')
		V(x-x')\psi_{v\sigma'}(x')+\hc.
	\end{equation}
	The fermionic fields can now be integrated out to to obtain 
	\begin{equation}
		\mathcal Z = \int \DD{\phi,\bar\phi,{\mathcal P},\bar{\mathcal P}}e^{-\mathcal S_{\mathrm{eff}}[\bar{\mathcal P},\mathcal P,\bar\phi,\phi]- S_{m}[\phi,\bar\phi]},
	\end{equation}
	with the effective action
	\begin{equation}
		\mathcal S_{\mathrm{eff}}=
		\int_{x,x'}\bar{\mathcal P}_{\sigma\sigma'}(x,x')V(x-x')
		\mathcal P_{\sigma'\sigma}(x',x)
		-\Tr\log[-\hat G_0^{-1}+\hat\Sigma^{\mathcal P}+\hat\Sigma^m],
	\end{equation}
	where hats denote matrices in band, spin, position, and imaginary-time space, and
	\begin{align}
		\hat\Sigma^{\mathcal P}_{\sigma\sigma'}(x,x')
		&=-V(x-x')\begin{pmatrix}
			0&\mathcal P_{\sigma\sigma'}(x,x')\\
			\bar{\mathcal P}_{\sigma'\sigma}(x',x)&0
		\end{pmatrix},\\
		\hat\Sigma^m_{\sigma\sigma'}(x)
		&=\begin{pmatrix}
			\Sigma^m_{cc;\sigma\sigma'}(x)&\Sigma'_{\sigma\sigma'}(x)\\
			[\Sigma'_{\sigma'\sigma}(x)]^*&\Sigma^m_{vv;\sigma\sigma'}(x)
		\end{pmatrix}.
	\end{align}
	We highlight here the off-diagonal blocks of the magnon self-energy, $\Sigma'_{\sigma\sigma'}(x)$.
	
	We now perform a saddle-point expansion around the mean-field value of the polarization field, $\langle\mathcal P\rangle=0$,\footnote{This is strictly speaking not valid once the excitons condense, but we choose to not describe this at the level of the Hubbard-Stratanovich transformation, instead deriving the
		Gross-Pitaevskii equations; see \cref{sec:bec}.} and associate the fluctuations around this saddle point with field describing Wannier excitons. We then obtain
	\begin{equation}
		\Tr\log[-\hat G_0^{-1}+\hat\Sigma^{\mathcal P}+\hat\Sigma^m]
		\simeq-\frac12\Tr[\hat G_c^0\hat\Sigma^{\mathcal P}_{cv}\hat G_v^0\hat\Sigma^{\mathcal P}_{vc}]
		-\frac12\Tr[\hat G_c^0\hat\Sigma^{\mathcal P}_{cv}\hat G_v^0\hat\Sigma^m_{vc}]
		-\frac12\Tr[\hat G_c^0\hat\Sigma^m_{cv}\hat G_v^0\hat\Sigma^{\mathcal P}_{vc}].
		\label{eq:trlog}
	\end{equation}
	The first term generates the exciton propagator; the final two generate the bilinear magnon--exciton coupling. Terms quadratic in $\Sigma^m$ only renormalize the electronic dispersion at this order, which we have therefore neglected.
	
	We introduce relative momentum $\bm k$ and center-of-mass (CoM) momentum $\bm q$ through
	\begin{equation}
		\mathcal P_{\sigma\sigma'}(\bm k,\bm q,\ii\Omega_n)=
		\frac1{\sqrt\beta\mathcal V}\int\d{\bm x}\d{\bm x'}\d{\tau}\,
		\mathcal P_{\sigma\sigma'}(\bm x,\bm x',\tau) e^{-\ii\bm k\cdot(\bm x-\bm x')}
		e^{-\ii\bm q\cdot(\gamma_c\bm x+\gamma_v\bm x')}
		e^{\ii\Omega_n\tau},
	\end{equation}
	where $M_X^{\sigma\sigma'}=m_c^\sigma+m_v^{\sigma'}$, $\gamma_c=m_c^\sigma/M_X^{\sigma\sigma'}$, and $\gamma_v=m_v^{\sigma'}/M_X^{\sigma\sigma'}$, so that $\gamma_c+\gamma_v=1$.\footnote{For simplicity, we assume the effective-mass approximation to hold. The results here are however generic, and can also be calculated for arbitrary bands.} The polarization self-energy is then
	\begin{equation}
		\Sigma^{\mathcal P}_{\sigma\sigma'}(\bm k,\bm q,\ii\Omega_n)
		=-\frac1{\mathcal V}\sum_{\bm p}V(\bm k-\bm p)
		\mathcal P_{\sigma\sigma'}(\bm p,\bm q,\ii\Omega_n).
	\end{equation}
	
	The quadratic polarization action thus becomes
	\begin{equation}
		\mathcal S_{\mathcal P}=-\sum_n\sum_{\bm q}\sum_{\bm k,\bm k'}
		\bar{\mathcal P}_{\sigma\sigma'}(\bm k,\bm q,\ii\Omega_n) G^{-1}_{0;\mathcal P;\sigma\sigma'}(\bm k,\bm k',\bm q,\ii\Omega_n)
		\mathcal P_{\sigma\sigma'}(\bm k',\bm q,\ii\Omega_n),
	\end{equation}
	with the polarization Green function given by
	\begin{equation}
		-\mathcal V G^{-1}_{0;\mathcal P;\sigma\sigma'}(\bm k,\bm k',\bm q,\ii\Omega_n)
		=V(\bm k-\bm k')\\
		+\frac1{\mathcal V}\sum_{\bm p}V(\bm k-\bm p)
		\Pi^{cv}_{\sigma\sigma'}(\bm p+\gamma_c\bm q,\bm p-\gamma_v\bm q,\ii\Omega_n)
		V(\bm p-\bm k').
	\end{equation}
	Here
	\begin{align}
		\Pi^{cv}_{\sigma\sigma'}(\bm k,\bm k',\ii\Omega_n)
		&=\frac1\beta\sum_mG_{c\sigma}(\bm k,\ii\Omega_n+\ii\omega_m)
		G_{v\sigma'}(\bm k',\ii\omega_m)\\
		&=\frac{N_F(\zeta^c_{\sigma\bm k})-N_F(\zeta^v_{\sigma'\bm k'})}
		{\zeta^c_{\sigma\bm k}-\zeta^v_{\sigma'\bm k'}-\ii\Omega_n}
	\end{align}
	is the generalized Lindhard function.
	
	As can be directly obtained from \cref{eq:trlog}, only the off-diagonal magnon self-energy $\Sigma'$ couples linearly to $\mathcal P$. 
	Using the Fourier convention
	\begin{equation}
		\Sigma'_{\sigma\sigma'}(\bm q,\ii\Omega_n)
		=\frac1{\sqrt{\beta\mathcal V}}\int\d{\bm x}\d{\tau}\,
		\Sigma'_{\sigma\sigma'}(\bm x,\tau)e^{-\ii\bm q\cdot\bm x}e^{\ii\Omega_n\tau},
	\end{equation}
	the coupling action can be written as
	\begin{equation}
		\mathcal S_c=-\frac1{\mathcal V\sqrt\beta}\sum_{n,\bm q,\bm k}
		\mathcal K_{\sigma\sigma'}(\bm k,\bm q,\ii\Omega_n)
		\left[\Sigma'_{\sigma\sigma'}(\bm q,\ii\Omega_n)
		\bar{\mathcal P}_{\sigma\sigma'}(\bm k,\bm q,\ii\Omega_n)+\hc\right],
	\end{equation}
	where we have defined the screened Lindhard function
	\begin{equation}
		\mathcal K_{\sigma\sigma'}(\bm k,\bm q,\ii\Omega_n)
		\equiv-\frac1{\mathcal V}\sum_{\bm p}V(\bm k-\bm p)
		\Pi^{cv}_{\sigma\sigma'}(\bm p+\gamma_c\bm q,\bm p-\gamma_v\bm q,\ii\Omega_n).
	\end{equation}
	\section{Exciton operators}
	
	The polarization action is diagonalized by solving $G^{-1}_{0;\mathcal P}\!\cdot\!\Phi=0$. At zero temperature, this is the eigenvalue problem known as the Bethe--Salpeter equation (BSE) \cite{combescotExcitonsCooperPairs2015},
	\begin{equation}
		\mathcal E^{cv}_{\sigma\sigma';\bm q\bm k}\Phi^{\sigma\sigma'}_{\nu\bm q}(\bm k)
		-\frac1{\mathcal V}\sum_{\bm k'}V(\bm k-\bm k')
		\Phi^{\sigma\sigma'}_{\nu\bm q}(\bm k')
		=\epsilon^\nu_{\bm q}\Phi^{\sigma\sigma'}_{\nu\bm q}(\bm k),
	\end{equation}
	where
	\begin{equation}
		\mathcal E^{cv}_{\sigma\sigma';\bm q\bm k}
		\equiv\zeta^c_{\sigma,\bm k+\gamma_c\bm q}
		-\zeta^v_{\sigma',\bm k-\gamma_v\bm q}.
	\end{equation}
	Here $\nu$ labels the relative-motion eigenstates. With the normalization $\mathcal V^{-1}\sum_{\bm k}|\Phi^{\sigma\sigma'}_{\nu\bm q}(\bm k)|^2=1$, we expand
	\begin{equation}
		\mathcal P_{\sigma\sigma'}(\bm k,\bm q,\ii\Omega_n)
		=\frac1{\sqrt{\mathcal V}}\sum_\nu
		\Phi^{\sigma\sigma'}_{\nu\bm q}(\bm k)X_{\nu;\bm q}(\ii\Omega_n).
	\end{equation}
	The polarization action becomes
	\begin{equation}
		\mathcal S_{\mathcal P}=-\sum_{n,\bm q,\nu}
		\bar X_{\nu;\bm q}(\ii\Omega_n)G^{-1}_{X;\nu}(\bm q,\ii\Omega_n)
		X_{\nu;\bm q}(\ii\Omega_n),
	\end{equation}
	where
	\begin{equation}
		G^{-1}_{X;\nu}(\bm q,\ii\Omega_n)=\ii\Omega_n-\epsilon^\nu_{\bm q}.
	\end{equation}
	The coupling action expressed in these exciton operators is now,
	\begin{equation}
		\mathcal S_c=\frac1{\sqrt{\beta\mathcal V}}\sum_{n,\bm q,\nu}
		\widetilde{\mathcal K}_{\nu;\sigma\sigma'}(\bm q,\ii\Omega_n)\\
		\left[\Sigma'_{\sigma\sigma'}(\bm q,\ii\Omega_n)
		\bar X_{\nu;\bm q}(\ii\Omega_n)+\hc\right],
		\label{eq:Sc}
	\end{equation}
	with
	\begin{equation}
		\widetilde{\mathcal K}_{\nu;\sigma\sigma'}(\bm q,\ii\Omega_n)
		\equiv-\int\frac{\d{\bm k}}{(2\pi)^2}\,
		\mathcal K_{\sigma\sigma'}(\bm k,\bm q,\ii\Omega_n)
		\Phi^{\sigma\sigma'}_{\nu\bm q}(\bm k).
	\end{equation}
	
	For parabolic bands,
	\begin{equation}
		\mathcal E^{cv}_{\sigma\sigma';\bm q\bm k}=E_g^{\sigma\sigma'}
		+\frac{q^2}{2M_X^{\sigma\sigma'}}+\frac{k^2}{2\mu_{\sigma\sigma'}},
	\end{equation}
	where $E_g^{\sigma\sigma'}$ is the spin-resolved band gap and $\mu_{\sigma\sigma'}=m_c^\sigma m_v^{\sigma'}/M_X^{\sigma\sigma'}$ is the reduced mass. With the two-dimensional Coulomb interaction $V(\bm k)=e^2/(2\epsilon k)$, with $\epsilon\equiv \epsilon_0\epsilon_r$, the exciton dispersion is
	\begin{equation}
		\epsilon^\nu_{\bm q}=\Delta_\nu^{\sigma\sigma'}
		+\frac{q^2}{2M_X^{\sigma\sigma'}},
	\end{equation}
	where $	\Delta_\nu^{\sigma\sigma'}\equiv E_g^{\sigma\sigma'}-E_B^\nu$ is the exciton gap.
	At zero temperature,
	\begin{equation}
		\mathcal K_{\sigma\sigma'}(\bm k,\bm q,\ii\Omega_n)
		=-\int\frac{\d{\bm p}}{(2\pi)^2}V(\bm k-\bm p)
		\widetilde G_{\sigma\sigma'}(\bm p,\bm q,\ii\Omega_n),
	\end{equation}
	where
	\begin{equation}
		\widetilde G_{\sigma\sigma'}(\bm p,\bm q,\ii\Omega_n)
		\equiv\frac{2\mu_{\sigma\sigma'}}{\kappa^2_{\sigma\sigma'}(\bm q,\ii\Omega_n)+p^2}
	\end{equation}
	is a Yukawa potential, with
	\begin{equation}
		\kappa^2_{\sigma\sigma'}(\bm q,\ii\Omega_n)
		\equiv2\mu_{\sigma\sigma'}\left(E_g^{\sigma\sigma'}
		+\frac{q^2}{2M_X^{\sigma\sigma'}}-\ii\Omega_n\right).
		\label{eq:kappa}
	\end{equation}
	characterizing the distance over which a virtual electron--hole excitation propagates coherently.
	
	The convolution can be expressed as
	\begin{equation}
		\mathcal K_{\sigma\sigma'}(\bm k,\bm q,\ii\Omega_n)
		=-\frac{1}{2a_{B;\sigma\sigma'}
			\sqrt{\kappa^2_{\sigma\sigma'}(\bm q,\ii\Omega_n)+k^2}}\\
		\mathrm K\!\left(\frac{k^2}{k^2+\kappa^2_{\sigma\sigma'}(\bm q,\ii\Omega_n)}\right),
	\end{equation}
	where $\mathrm K(m)=\int_0^{\pi/2}\d{\theta}/\sqrt{1-m\sin^2\theta}$ is the complete elliptic integral of the first kind and
	\begin{equation}
		a_{B;\sigma\sigma'}\equiv\frac{4\pi\epsilon}{\mu_{\sigma\sigma'}e^2}.
	\end{equation}
	is the exciton Bohr radius.
	
	This concludes the formal derivatoin. To make further progress, we have to specify both the exciton wavefunction $\Phi_{\nu\bm q}^{\sigma\sigma'}(\bm k)$ and the the off-diagonal components of the magnon self-energy $\Sigma'_{\sigma\sigma'}(\bm q,\ii\Omega_n)$. We focus here on the $1s$-exciton, since this is typically the lowest energy solution of the BSE. This will allow to perform some additional calculations before having to specify $\Sigma'_{\sigma\sigma'}(\bm q,\ii\Omega_n)$.
	\subsection{\texorpdfstring{$1s$}{1s}-exciton--magnon overlap}
	We restrict ourselves now to the lowest two-dimensional $1s$ state, where
	\begin{equation}
		\Phi_{1s}(\bm k)=\frac{\sqrt{8\pi}\,a_{1s}}{[1+(ka_{1s})^2]^{3/2}},
	\end{equation}
	with $a_{1s}={a_B}/{2}=1/{\sqrt{2\mu E_B}}$.
	Here we have neglected the CoM dependence of the relative wave function in the long-wavelength regime. This allows us to rewrite $\kappa$, \cref{eq:kappa}, in a particularly transparent form,
	\begin{equation}
		\kappa^2(\bm q,\ii\Omega_n)=a_{1s}^{-2}
		+2\mu\left(\epsilon_{\bm q}-\ii\Omega_n\right),
	\end{equation}
	where $ \epsilon_{\bm q}=\Delta_X+\frac{q^2}{2M_X}$ is the exciton dispersion.
	Writing $z\equiv a_{1s}\kappa(\bm q,\ii\Omega_n)$, the overlap in \cref{eq:Sc} is
	\begin{equation}
		\widetilde{\mathcal K}_{1s}(\bm q,\ii\Omega_n)
		=-\sqrt{\frac{\pi}{8}}\frac{F(z)}{a_{1s}^2\kappa(\bm q,\ii\Omega_n)},
		\label{eq:overlap}
	\end{equation}
	where
	\begin{equation}
		F(z)\equiv\frac{2z}{\pi}\int_0^\infty\d{u}\,
		\frac{\mathrm K[u^2/(u^2+z^2)]}{\sqrt{u^2+z^2}\,(1+u^2)^{3/2}}.
	\end{equation}
	The form factor satisfies $F(z)\to1$ for $z\gg1$. On the exciton pole, however, $\ii\Omega_n\to\epsilon_{\bm q}$ and hence $z=1$; numerical evaluation gives $F(1)\simeq0.886$. We thus find
	\begin{equation}
		\widetilde{\mathcal K}_{1s}(\bm q,\ii\Omega_n)
		\simeq -F(1)\sqrt{\frac{\pi}{8}}\frac1{a_{1s}^2\kappa(\bm q,\ii\Omega_n)}.
	\end{equation}
	This concludes the analysis for the $1s$-excitons, and we next need to need to specify the off-diagonal magnon self-energy, which
	is model dependent.
	\section{Ferromagnetic van der Waals bilayer}
	
	\subsection{Magnon dynamics}
	For completeness, we repeat here the derivation of the magnon dynamics, starting from the same minimal Heisenberg model as in the End Matter of the main text,
	\begin{equation}
		H_m=-\frac12\sum_{\gamma\in\{t,b\}}\Biggl[
		\sum_{\langle ij\rangle}J^\parallel\bm S_{\gamma,i}\cdot\bm S_{\gamma,j}
		+\sum_iJ^\perp\bm S_{\gamma,i}\cdot\bm S_{\bar\gamma,i}\\
		+K\sum_i(S^z_{\gamma,i})^2\Biggr].
	\end{equation}
	Here $J^\parallel>0$ and $J^\perp>0$ are the intra- and interlayer exchange couplings, and $K>0$ is the easy-axis anisotropy. Applying the Holstein--Primakoff transformation, $S^+_{\gamma,i}\simeq\sqrt{2S}\phi_{\gamma,i}$, and Fourier transforming gives
	\begin{equation}
		H_m=\sum_{\gamma,\bm k}\left[(\Delta_a+\rho_sk^2+J')
		\phi^\dagger_{\gamma;\bm k}\phi_{\gamma;\bm k}
		-J'\phi^\dagger_{\gamma;\bm k}\phi_{\bar\gamma;\bm k}\right],
	\end{equation}
	where
	\begin{equation}
		\Delta_a=SK,\qquad \rho_s=J^\parallel Sa^2,\qquad J'=J^\perp S.
	\end{equation}
	The acoustic and optical fields are
	\begin{equation}
		\phi_{a;\bm k}=\frac{\phi_{t;\bm k}+\phi_{b;\bm k}}{\sqrt2},
		\qquad
		\phi_{o;\bm k}=\frac{\phi_{t;\bm k}-\phi_{b;\bm k}}{\sqrt2},
	\end{equation}
	and their dispersions are
	\begin{equation}
		\omega^a_{\bm k}=\Delta_a+\rho_sk^2,
		\qquad
		\omega^o_{\bm k}=\Delta_o+\rho_sk^2,
		\qquad \Delta_o=\Delta_a+2J'.
	\end{equation}
	
	\subsection{Electron Hamiltonian and transverse exchange}
	The electron Hamiltonian is
	\begin{equation}
		H_e=\sum_{\bm k}\sum_{\substack{\sigma\sigma'\\
				\tau\tau'\in\{c,v\}\\\gamma\gamma'\in\{t,b\}}}
		d^\dagger_{\gamma\tau\sigma;\bm k}\\
		\left[\tau_z\left(\frac{E_g}{2}+\xi_{\bm k}\right)
		+\tau_x\gamma_xt+\gamma_z\frac V2 -\sigma_z I_{sd}^{\gamma\tau}S\right]
		d_{\gamma'\tau'\sigma';\bm k},
	\end{equation}
	where $\xi_{\bm k}=k^2/(2m)$. We have allowed for a finite tunneling element between the top and bottom layer, parameterized by $t$, which will be key to allow for an interband magnon self-energy. The voltage bias $V\equiv V_t-V_b$ allows for control over the bandgap, bringing the bands in the top and bottom layer close to each other. The static coupling to the local magnetic moments is described by an effective $sd$-like coupling, $I_{sd}^{\gamma\tau}$, which we allow to be layer- and band-dependent. 
	The Pauli matrices $\gamma_i$, $\tau_i$, and $\sigma_i$ act in layer, conduction/valence-band, and spin space, respectively. We retain the bottom-layer conduction band and top-layer valence band. In this two-band subspace,
	\begin{equation}
		H_e=\sum_{\sigma,\bm k}
		\begin{pmatrix}d^\dagger_{b,c,\sigma;\bm k}&d^\dagger_{t,v,\sigma;\bm k}\end{pmatrix}
		\begin{pmatrix}
			A_{\bm k}-\sigma I^{bc}_{sd}S&t\\
			t&-A_{\bm k}-\sigma I^{tv}_{sd}S
		\end{pmatrix}
		\begin{pmatrix}d_{b,c,\sigma;\bm k}\\d_{t,v,\sigma;\bm k}\end{pmatrix},
	\end{equation}
	where $A_{\bm k}=\xi_{\bm k}+E_g/2-V/2$. Defining
	\begin{equation}
		E_{\sigma\bm k}=\sqrt{t^2+(A_{\bm k}-\sigma I_{sd}^z)^2},
		\qquad \tan r_{\sigma \bm k}=\frac{t}{A_{\bm k}-\sigma I_{sd}^z},
	\end{equation}
	the band operators are
	\begin{align}
		c_{\sigma\bm k}&=\cos\frac{r_{\sigma \bm k}}2d_{b,c,\sigma;\bm k}
		+\sin\frac{r_{\sigma \bm k}}2d_{t,v,\sigma;\bm k},\\
		v_{\sigma\bm k}&=-\sin\frac{r_{\sigma \bm k}}2d_{b,c,\sigma;\bm k}
		+\cos\frac{r_{\sigma \bm k}}2d_{t,v,\sigma;\bm k},
	\end{align}
	and
	\begin{equation}
		H_e=\sum_{\sigma,\bm k}\left[
		c^\dagger_{\sigma\bm k}\left(E_{\sigma\bm k}-\sigma I_{sd}^0S\right)c_{\sigma\bm k}
		-v^\dagger_{\sigma\bm k}\left(E_{\sigma\bm k}-\sigma I_{sd}^0S\right)v_{\sigma\bm k}
		\right],
	\end{equation}
	where $I_{sd}^{0/z}=({I_{sd}^{bc}\pm I_{sd}^{tv}})/2$.
	
	For $I_{sd}^{0}<0$, the  lowest conduction and highest valence bands are now $c_\downarrow$ and $v_\uparrow$. Expanding at small momentum gives
	\begin{equation}
		H_e=\sum_{\bm k}\left[
		c^\dagger_{\downarrow\bm k}\left(\frac\Delta2+\frac{k^2}{2m}\right)c_{\downarrow\bm k}
		-v^\dagger_{\uparrow\bm k}\left(\frac\Delta2+\frac{k^2}{2m}\right)v_{\uparrow\bm k}
		\right],
	\end{equation}
	where
	\begin{align}
		\Delta&=\sqrt{t^2+(E_g/2-V/2+I_{sd}^z S)^2}+\sqrt{t^2+(E_g/2-V/2-I_{sd}^z S)^2}-2|I_{sd}^0|S\\
		&\approx2\sqrt{t^2+(E_g/2-V/2)^2}-2|I_{sd}^0|S
	\end{align}
	and we have neglected corrections to the effective mass, which are of the order $O(t^2/(E_g/2-V/2)^2)$, and neglected the corrections of the band gap due to  $I_{sd}^zS$, which are of order $(I_{sd}^zS)^2/(E_g/2-V/2)^2$.
	
	The local $sd$ exchange interaction is
	\begin{equation}
		H_{sd}=-\frac1{\sqrt{\mathcal N}}
		\sum_{\substack{\sigma\sigma'\\\tau\in\{c,v\}\\\gamma\in\{t,b\}}}
		\sum_{\bm q,\bm k}I_{sd}^{\gamma\tau}\bm S_{\gamma;\bm q}\cdot
		d^\dagger_{\gamma\tau\sigma;\bm k+\bm q}\bm\sigma_{\sigma\sigma'}
		d_{\gamma\tau\sigma';\bm k}.
	\end{equation}
	Keeping only interband transverse terms in the $c_\downarrow,v_\uparrow$ subspace gives
	\begin{equation}
		H_{sd}=\frac1{\sqrt{\mathcal N}}\sum_{\bm q,\bm k}
		\left[c^\dagger_{\downarrow\bm k+\bm q}W[\bm S_{\bm q}]v_{\uparrow\bm k}+\hc\right],
	\end{equation}
	where
	\begin{equation}
		W[\bm S_{\bm q}]=-\frac1{\sqrt2}\frac{t}{\sqrt{t^2+(E_g/2-V/2)^2}}
		\left(I_{sd}^zS^+_{a;\bm q}+I_{sd}^0S^+_{o;\bm q}\right).
	\end{equation}
	We have defined
	\begin{equation}
		\sqrt2\bm S_{a/o;\bm q}=\bm S_{t;\bm q}\pm\bm S_{b;\bm q}.
	\end{equation}
	Thus, $I_{sd}^{bc}=I_{sd}^{tv}$ implies $I_{sd}^z=0$ and suppresses the acoustic bilinear coupling, whereas $I_{sd}^{bc}=-I_{sd}^{tv}$ implies $I_{sd}^0=0$ and suppresses the optical bilinear coupling.
	Using $S^+_{\eta;\bm q}=\sqrt{2S}\phi_{\eta;\bm q}$, the vertex becomes
	\begin{equation}
		W[\bm S_{\bm q}]=\sqrt S\sum_{\eta\in\{a,o\}}g_\eta\phi_{\eta;\bm q},
	\end{equation}
	with
	\begin{equation}
		g_\eta=-\frac{t}{\sqrt{t^2+(E_g/2-V/2)^2}}
		\begin{cases}
			I_{sd}^z,&\eta=a,\\
			I_{sd}^0,&\eta=o.
		\end{cases}
	\end{equation}
	Consistent with the local approximation, we have neglected $\bm k$ dependency of the interaction vertex, $W_{\bm k}[\bm S_{\bm q}]\approx W[\bm S_{\bm q}]$. This amounts to the approximation that the exciton wavefunction is localized such that $\varepsilon_{\bm k}<E_g/2-V/2$ for the relevant internal exciton momenta $\bm k$. 
	
	We thus conclude that within such a model system, the tunneling enables a coupling of the magnons to transitions between the conduction and valence band. 
	
	\subsection{Bilinear magnon--exciton coupling}
	
	Promoting the magnon operators to continuum fields gives the off-diagonal magnon self-energy,
	\begin{equation}
		\Sigma'_{\downarrow\uparrow}(\bm q,\ii\Omega_n)
		=-\sqrt{\beta\mathcal V}\,l_m\sum_{\eta\in\{a,o\}}g_\eta
		\phi_{\eta;\bm q}(\ii\Omega_n),
	\end{equation}
	where
	\begin{equation}
		l_m=\sqrt{A_{\mathrm{uc}}S}
	\end{equation}
	is the magnon normalization length and $A_{\mathrm{uc}}$ is the unit-cell area. We now obtain the coupling action as
	\begin{equation}
		\mathcal S_c=\sum_{n,\bm q}\sum_{\eta\in\{a,o\}}
		J^\eta_{\bm q}(\ii\Omega_n)
		\left[\bar X_{\bm q}(\ii\Omega_n)\phi_{\eta;\bm q}(\ii\Omega_n)+\hc\right],
	\end{equation}
	where
	\begin{equation}
		J^\eta_{\bm q}(\ii\Omega_n)=g_\eta l_m
		\widetilde{\mathcal K}_{1s}(\bm q,\ii\Omega_n).
	\end{equation}
	This thus becomes the expression used in the main text,
	\begin{equation}
		J^\eta_{\bm q}(\ii\Omega_n)=g_\eta F(1) \sqrt{\frac\pi8}\frac{l_m}{a_{1s}}
		\sqrt{\frac{E_B}{\Delta+q^2/(2M_X)-\ii\Omega_n}}.
	\end{equation}
	
	The complete quadratic action can therefore be written as
	\begin{equation}
		\mathcal S=\sum_{n,\bm q}\vec\Psi^\dagger_{\bm q}(\ii\Omega_n)
		\begin{pmatrix}
			-G_X^{-1}&J^a_{\bm q}(\ii\Omega_n)&J^o_{\bm q}(\ii\Omega_n)\\
			J^a_{\bm q}(\ii\Omega_n)&-G_a^{-1}&0\\
			J^o_{\bm q}(\ii\Omega_n)&0&-G_o^{-1}
		\end{pmatrix}
		\vec\Psi_{\bm q}(\ii\Omega_n),
		\label{eq:full-S}
	\end{equation}
	where $\vec\Psi_{\bm q}(\ii\Omega_n)=(X_{\bm q}(\ii\Omega_n),\phi_{a;\bm q}(\ii\Omega_n),\phi_{o;\bm q}(\ii\Omega_n))^T$, $G_X^{-1}=\ii\Omega_n-\epsilon_{\bm q}$, and $G_{a/o}^{-1}=\ii\Omega_n-\omega^{a/o}_{\bm q}$. In the on-shell approximation, $\ii\Omega_n\to\epsilon_{\bm q}$, the coupling becomes momentum independent:
	\begin{equation}
		J_\eta=g_\eta F(1)\sqrt{\frac\pi8}\frac{l_m}{a_{1s}}.
	\end{equation}
	The corresponding Hamiltonian is
	\begin{equation}
		H_{Xm}=\sum_{\bm q}\vec\Psi^\dagger_{\bm q}
		\begin{pmatrix}
			\epsilon_{\bm q}&J_a&J_o\\
			J_a&\omega^a_{\bm q}&0\\
			J_o&0&\omega^o_{\bm q}
		\end{pmatrix}
		\vec\Psi_{\bm q}.
	\end{equation}
	Here $\vec\Psi_{\bm q}=(X_{\bm q},\phi_{a;\bm q},\phi_{o;\bm q})^T$ contains the exciton and magnon annihilation operators.
	
\begin{figure}
		\centering
		\includegraphics[width=0.8\textwidth]{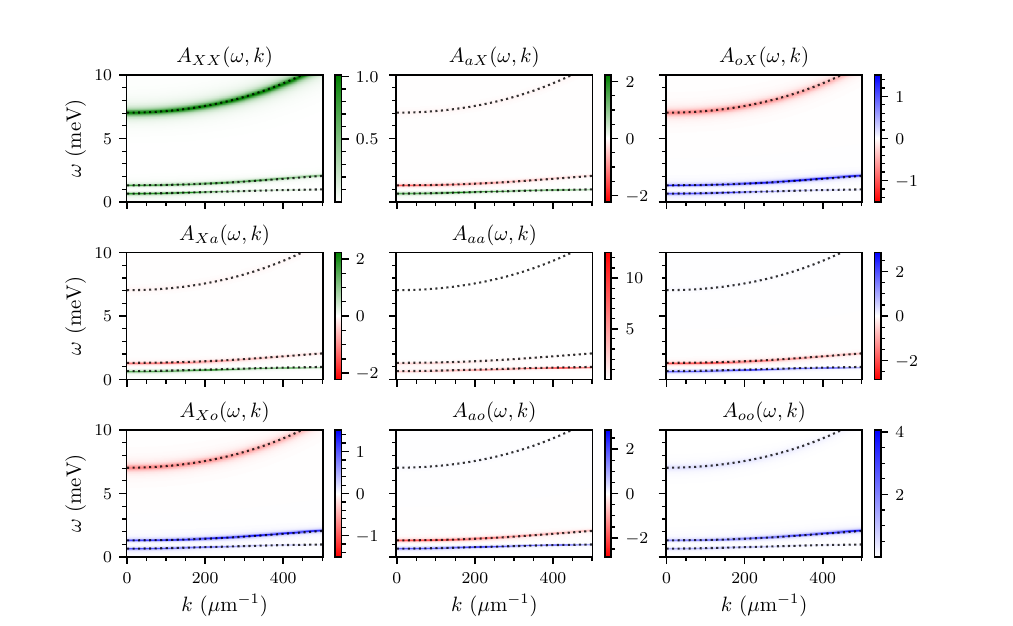}
		\caption{Spectral function of the full frequency-dependent Green function. The dotted curves show the dispersions obtained from the on-shell Hamiltonian approximation. Here $X$, $a$, and $o$ denote the exciton, acoustic-magnon, and optical-magnon components, respectively.}
		\label{fig:spectral-sm}
	\end{figure}

	After analytic continuation, direct inversion of the matrix in \cref{eq:full-S} gives the spectral function
	\begin{equation}
		A_{\alpha\beta}(\omega,k)=-\frac1\pi\Imm[G_{\alpha\beta}(\omega,k)].
	\end{equation}
	In \cref{fig:spectral-sm}, we compare its resonances with the eigenenergies of the on-shell Hamiltonian approximation. These results demonstrate that for our model system, the Hamiltonian approximation is accurate. In what follows, we thus continue with the Hamiltonian approximation.
	
	\subsection{Longitudinal exchange and density coupling}
	
	To leading order in tunneling, the longitudinal exchange projected onto the two low-energy bands is
	\begin{equation}
		H_{sd}^z=\frac1{\sqrt{2\mathcal N}}\sum_{\bm k,\bm q}
		\left(I_{sd}^{bc}(S^z_{a;\bm q}-S^z_{o;\bm q}) c^\dagger_{\downarrow\bm k+\bm q}c_{\downarrow\bm k}
		-I_{sd}^{tv}(S^z_{a;\bm q}+S^z_{o;\bm q})v^\dagger_{\uparrow\bm k+\bm q}v_{\uparrow\bm k}\right)+O(t),
	\end{equation}
	such that the relevant magnon self-energy is
	\begin{equation}
		\Sigma^m_{cc;\sigma\sigma'}(\bm q,\ii\Omega_n)=\delta_{\sigma,\sigma'}\delta_{\sigma,\downarrow}\sum_{\eta\in\{a,o\}}\gamma_{\eta;c}S^z_{\eta;\bm q}(\ii\Omega_n);\quad \Sigma^m_{vv;\sigma\sigma'}(\bm q,\ii\Omega_n)=\delta_{\sigma,\sigma'}\delta_{\sigma,\uparrow}\sum_{\eta\in\{a,o\}}\gamma_{\eta;v}S^z_{\eta;\bm q}(\ii\Omega_n),
	\end{equation}
	where
	\begin{equation}
		\gamma_{a;c}=\frac{I_{sd}^0+I_{sd}^z}{\sqrt2},\quad
		\gamma_{a;v}=-\frac{I_{sd}^0-I_{sd}^z}{\sqrt2},\quad
		\gamma_{o;c}=-\frac{I_{sd}^0+I_{sd}^z}{\sqrt2},\quad
		\gamma_{o;v}=-\frac{I_{sd}^0-I_{sd}^z}{\sqrt2}.
	\end{equation}
	We thus obtain the first contribution in the cubic expansion of the $\Tr\log$ in \cref{eq:trlog} as
	\begin{equation}
		\mathcal S_3=\Tr[ \hat \Sigma^{m}_{cc} \hat G_{c}^0  \hat \Sigma^{\mathcal P}_{cv} \hat G_v^0 \hat\Sigma^{\mathcal P}_{vc} \hat G_c^0]  +\Tr[ \hat \Sigma^{m}_{vv} \hat G_{v}^0  \hat \Sigma^{\mathcal P}_{vc} \hat G_c^0 \hat\Sigma^{\mathcal P}_{cv} \hat G_v^0].
	\end{equation}
	Our main reason to isolate this interaction is to study its impact on the Bose-Einstein condensation of the combined magnon-exciton system. We will thus focus on the equal-time and equal-space interaction, i.e., on terms that will results in contributions to the action, $\sim S_z(\bm x,\tau)\bar X(\bm x,\tau)X(\bm x,\tau)$. This interaction will subsequently enter the Gross-Pitaeviskii equations governing the condensate wavefunction.

	Performing the trace and Fourier transforming, we then obtain
	\begin{equation}
		\mathcal S_3=
		\frac{1}{\mathcal V^2}
		\sum_{\eta\in\{a,o\}}
		\sum_{n,m}
		\sum_{\bm p,\bm q}
		\sum_{\bm k,\bm k'}
		\mathcal G_{\eta;\bm k\bm k'}\,
		S^z_{\eta;\bm p}(\ii\Omega_m)
		\mathcal P_{\downarrow\uparrow}
		(\bm k,\bm q,\ii\Omega_n)
		\bar{\mathcal P}_{\downarrow\uparrow}
		(-\bm k',\bm q+\bm p,\ii\Omega_{n+m}),
	\end{equation}
	where we have focused only on the equal-space and equal-time interaction and thus have a static kernel
	\begin{equation}
		\mathcal G_{\eta;\bm k\bm k'}
		=
		(\gamma_{\eta;c}-\gamma_{\eta;v})
		\int\frac{\d{\bm\ell}}{(2\pi)^2}
		\frac{
			V(\bm k-\bm\ell)V(\bm\ell-\bm k')
		}{
			\left(
			\zeta^c_{\downarrow,\bm\ell}
			-\zeta^v_{\uparrow,\bm\ell}
			\right)^2
		}.
	\end{equation}
	Expanding the polarization fields $\mathcal P$ in the exciton operators, we find,
	\begin{equation}
		\mathcal S_3=
		\sum_{\eta\in\{a,o\}}
		\sum_{n,m}
		\sum_{\bm p,\bm q}
		\tilde\Gamma_\eta\,
		S^z_{\eta;\bm p}(\ii\Omega_m)
		X_{\bm q}(\ii\Omega_n)
		\bar X_{\bm q+\bm p}(\ii\Omega_{n+m}),
	\end{equation}
	where 
	\begin{equation}
		\tilde\Gamma_\eta=
		(\gamma_{\eta;c}-\gamma_{\eta;v})
		\int\frac{\d{\bm\ell}}{(2\pi)^2}
		\frac{|C(\bm\ell)|^2}{
			\left(
			\zeta^c_{\downarrow,\bm\ell}
			-\zeta^v_{\uparrow,\bm\ell}
			\right)^2
		},
	\end{equation}
	and \begin{align}
		C(\bm\ell)&=
		\int\frac{\d{\bm k}}{(2\pi)^2}
		V(\bm k-\bm\ell)\Phi_{1s}(\bm k)\\
		&=\int \frac{\d{\bm k}}{(2\pi)^2} \frac{V(\bm k-\bm \ell) \sqrt{8\pi}a_{1s}}{(1+(ka_{1s})^2)^{3/2}},
	\end{align}
	where the function $C(\bm \ell)$ is most conveniently calculated as a convolution 
	\begin{equation}
		C(\bm \ell)= \sqrt{\frac{2}{\pi}}\frac{e^2}{4\pi \epsilon a_{1s}} \int {\d{x}\d\theta}  e^{-x/a_{1s}-\ii \ell x\cos\theta} = \sqrt{8\pi}\frac{e^2}{4\pi \epsilon}\frac{1}{\sqrt{1+a_{1s}^2 \ell^2}}.
	\end{equation}
	We thus have the vertex
	\begin{equation}
		\tilde\Gamma_\eta = 4(\gamma_{\eta;c}-\gamma_{\eta;v})\left(\frac{e^2}{4\pi \epsilon}\right)^2 \frac{1}{2a_{1s}^2\Delta ^2} \frac{b-1-\log{b}}{(b-1)^2};\quad b\equiv \frac{1}{2\mu \Delta a_{1s}^2} = \frac{E_B}{\Delta } \label{eq:Gamma-full}
	\end{equation}
	where close to  we have that $b\approx 1$, and thus 
	\begin{equation}
		\tilde\Gamma_\eta = \frac{\gamma_{\eta;c}-\gamma_{\eta;v}}{a_{1s}^2\Delta^2}\left(\frac{e^2}{4\pi \epsilon}\right)^2 +O(b-1)
	\end{equation}
	which can be further written in terms of the binding energy using the relation $\frac{e^2}{4\pi \epsilon}a_{1s}=\frac{1}{2\mu}$, such that
	\begin{equation}
		\tilde\Gamma_\eta = (\gamma_{\eta;c}-\gamma_{\eta;v})\frac{ E_B^2}{\Delta^2} +O(b-1). 
	\end{equation}
	Fourier transforming the action back now to time and space, and expanding $\sqrt{2}S^z_a(\bm x,\tau)=2S-\bar\phi_a(\bm x,\tau)\phi_a(\bm x,\tau)-\bar\phi_o(\bm x,\tau)\phi_o(\bm x,\tau)$ and $\sqrt{2}S^z_o(\bm x,\tau)=-\bar\phi_a(\bm x,\tau)\phi_o(\bm x,\tau)-\bar\phi_o(\bm x,\tau)\phi_a(\bm x,\tau)$ in magnon fields, we have that the final two-magnon--exciton-density interaction reads\footnote{Neglecting static shifts of the exciton gap.}
	\begin{equation}
		\mathcal  S_3 =  \int \d{ \bm x} \d \tau n_X(\bm x,\tau)\left[\Gamma_0\left(n_a(\bm x,\tau) + n_o(\bm x,\tau)\right)-\Gamma_z\left(\bar\phi_{a}(\bm x,\tau)\phi_o(\bm x,\tau)+\hc \right)\right] 
	\end{equation}
	where
	\begin{equation}
		\Gamma_{{0/z}} \equiv -\sqrt{2}A_{\mathrm{uc}} \frac{ E_b^2}{\Delta^2}I_{sd}^{0/z} \label{eq:Gamma0z}
	\end{equation}
	We note finally that for $I_sd^{0}<0$ we have $\Gamma_{{0/z}}>0$ and the density-density interaction is thus repulsive. We furthermore stress that the interaction is not perturbative in the tunneling element $t$, and can thus be a sizable effect. In the main text, we continue in the limit of small $I^z_{sd}$, such that $I_z\ll I_0$ and its effect can be neglected on the condensate. Finally, in the numerical evaluation we use the full expression for $\Gamma_0$, \cref{eq:Gamma-full}, and not its simplified form \cref{eq:Gamma0z}.
	\section{Bose--Einstein condensate}
	\label{sec:bec}
	
	We consider now Bose-Einstein condensation (BEC) of the combined exciton-magnon system. 	
	To describe an exciton condensate, one requires a quartic exciton-exciton interaction. Instead of deriving this from the expansion of the $\Tr\log$ in \cref{eq:trlog},\footnote{Deriving a quartic exciton-exciton interaction from the expansion of the $\Tr\log$ is a formidable task. See also Ref.~\cite{noordmanVariationalFieldtheoreticalApproach2025}.} we choose here to postulate a local-space and time exciton-exciton interaction, which is valid in the dilute regime \cite{golomedovCompositeBosonDescription2017}. Given the success of the Hamiltonian approximation, we also assume the Hamiltonian coupling $J_\eta$. We neglect the effect of $\Gamma_z$, which will couple the exciton density to the optical--acoustic hybridization. As such, it will induce a small mixing of the two modes, which we expect to not affect the main condensate characteristics.
	
	The final action thus becomes in Euclidean time and real space,
	\begin{align}
		\mathcal S=\int_x\Bigg\{&
		\bar X\left(\partial_\tau+\Delta_X-\frac{\nabla^2}{2M_X}\right)X
		+\frac g2n_X^2
		\nonumber\\
		&+\sum_{\eta\in\{a,o\}}\left[
		\bar\phi_\eta\left(\partial_\tau+\Delta_\eta-\rho_s\nabla^2\right)\phi_\eta
		+\Gamma_0 n_Xn_\eta
		+\left(\bar XJ_\eta\phi_\eta+\hc\right)
		\right]\Bigg\}.
		\label{eq:bec-action}
	\end{align}
	Here we use the shorthand notation $x\equiv(\bm x,\tau)$,  $n_{X}(x)\equiv \bar X(x) X(x)$ is the exciton density, $n_{a/o}(x)\equiv \bar \phi_{a/o}(x) \phi_{a/o}(x)$ is the magnon density.
	
	We expand about homogeneous, time-independent condensates,
	\begin{equation}
		X=X_0+X',\qquad \phi_{a/o}=\phi_{a/o;0}+\phi'_{a/o}.
	\end{equation}
	Requiring the linear fluctuation terms to vanish gives the coupled Gross--Pitaevskii equations
	\begin{align}
		(\Delta_X+gn_X+\Gamma_0n_a+\Gamma_0n_o)X_0
		+J_a\phi_{a;0}+J_o\phi_{o;0}&=0,\\
		(\Delta_a+\Gamma_0n_X)\phi_{a;0}+J_aX_0&=0,\\
		(\Delta_o+\Gamma_0n_X)\phi_{o;0}+J_oX_0&=0.
	\end{align}
	
	It is instructive to first solve for $I_{sd}^z=0$, such that $J_a=0$. We can then find in the limit of small $\Gamma_0$,
	\begin{equation}
		n_X=\frac{J_o^2/\Delta_o-\Delta_X}{g}\left[1-\frac{2J_o^2}{\Delta_o^2}\frac{\Gamma_0}{g}\right]+O(\Gamma_0^2/g^2),
		\qquad n_a=0,
		\qquad n_o=\frac{J_o^2}{\Delta_o^2}\left[1-\frac{2(J_o^2/\Delta_o-\Delta_X)}{\Delta_o}\frac{\Gamma_0}{g}\right]n_X+O(\Gamma_0^2/g^2).
	\end{equation}
	This thus corresponds to the condensation of the hybrid exciton--optical-magnon quasiparticles, with the new gap $\Delta_x - J_0^2/\Delta_o$. The exciton--magnon density--density interaction introduces corrections to the magnon and exciton densities, which are small in $\Gamma_0/g$.

	In the general case, where all couplings are non-zero, we have the relations
	\begin{equation}
		n_{a/o}=\frac{J_{a/o}^2}{(\Delta_{a/o}+\Gamma_{0}n_X)^2}n_X,
	\end{equation}
	and $n_X$ is thus the physical root of
	\begin{equation}
		\Delta_X+gn_X-
		\sum_{\eta\in\{a,o\}}\frac{J_\eta^2\Delta_\eta}
		{(\Delta_\eta+\Gamma_0 n_X)^2}=0.
		\label{eq:root-X0}
	\end{equation}
	The branch continuously connected to the stable dilute condensate satisfies
	$\sign(X_0/\phi_{\eta;0})=-\sign(J_\eta)$. Stability additionally requires the derivative of the left-hand side of \cref{eq:root-X0} to be positive:
	\begin{equation}
		g-\sum_{\eta\in\{a,o\}}
		\frac{2J_\eta^2\Delta_\eta\Gamma_0}
		{(\Delta_\eta+\Gamma_0 n_X)^3}>0.
	\end{equation}
	Through this procedure we identify the physical $n_X$.
	The condensation threshold is furthermore
	\begin{equation}
		\Delta_{X,c}=\frac{J_a^2}{\Delta_a}+\frac{J_o^2}{\Delta_o}.
	\end{equation}
	
	Expanding the action further to quadratic in fluctuations gives the collective excitations,
	\begin{equation}
		\mathcal S^{(2)}=-\sum_{n,\bm k}
		\vec\psi^\dagger(k)\hat G_B^{-1}(k)\vec\psi(k),
	\end{equation}
	where $k=(\bm k,\ii\Omega_n)$ and
	\begin{equation}
		\vec\psi(k)=
		\left(X',\bar X',\phi'_a,\bar\phi'_a,\phi'_o,\bar\phi'_o\right)^T.
	\end{equation}
	Here
	\begin{equation}
		-\hat G_B^{-1}(k)=
		\begin{pmatrix}
			-\hat G_X^{-1}(k)&\hat T_a(k)&\hat T_o(k)\\
			\hat T_a(k)&-\hat G_a^{-1}(k)&0\\
			\hat T_o(k)&0&-\hat G_o^{-1}(k)
		\end{pmatrix},
	\end{equation}
	with
	\begin{align}
		-\hat G_X^{-1}(k)&=-\ii\Omega_n\hat\sigma_z
		+\left(\Delta_X+\frac{k^2}{2M_X}+2gn_X+\Gamma_0n_a+\Gamma_0n_o\right)\hat\sigma_0
		+gn_X\hat\sigma_x,\\
		-\hat G_{a/o}^{-1}(k)&=-\ii\Omega_n\hat\sigma_z
		+\left(\omega^{a/o}_{\bm k}+\Gamma_{0}n_X\right)\hat\sigma_0,\\
		\hat T_{a/o}(k)&=J_{a/o}\hat\sigma_0
		+\Gamma_{0}\sqrt{n_Xn_{a/o}}(\hat\sigma_0+\hat\sigma_x).
	\end{align}
	
	The excitation energies follow from paraunitary diagonalization \cite{colpaDiagonalizationQuadraticBoson1978} of the corresponding Bogoliubov Hamiltonian. In particular,
	\begin{equation}
		-\hat G_B^{-1}(k)=-\ii\Omega_n\hat\tau_z+\hat H_B(\bm k),
		\qquad \hat\tau_z=\diag(1,-1,1,-1,1,-1).
	\end{equation}
	
	For $J_a=0$, the acoustic branch decouples from the exciton--optical-magnon subsystem at quadratic order. In the long-wavelength limit, and for $\rho_sM_X\ll1$, we find
	\begin{equation}
		E_+=\frac1{\Delta_o}\sqrt{(J_o^2+\Delta_o^2)^2+2J_o^2\Delta_ogn_X},
		\qquad
		E_-=\frac{\Delta_o}{E_+}\,ck,
	\end{equation}
	where
	\begin{equation}
		c=\sqrt{\frac{gn_X}{M_X}}
	\end{equation}
	is the bare exciton-condensate sound velocity. The acoustic-magnon gap is furthermore shifted to $\Delta_a+\Gamma_0n_X$.
	\bibliography{coupling}